# Entangled-path interferometer simpler, faster than LISA

P. B. Lerner[1]

**Abstract**

Entangled light can provide a seminal improvement in resolution sensitivity even without achieving Heisenberg limit in a single channel. In this paper, based on the "back-of-the-envelope" type calculations, I demonstrate an alternative path to space based long-arm interferometer. Its advantage with respect to LISA is that it does not require complex satellites with many active components to achieve similar resolution.

## Introduction

Demonstration of the gravitational waves in the Earth-based interferometers [1] became one of the pinnacles of more than century of optics research. Currently, LISA, a giant space-based interferometer is in its planning stages with the launch planned for 2030. [2] The author proposes much less ambitious scheme, which, if feasible, can be realized in less time and with less expense as an intermediate stage to LISA. This scheme is based on using entangled states and does not involve space launch and maintenance of high-powered lasers. The optical components on board of the satellites in the proposed scheme could be entirely passive.

Conventional proposals to use entangled light for improvement of the resolution of interferometers are based on achieving so-called Heisenberg limit $\Delta\phi_{res} = \frac{1}{N}$ where N is a number of photons in an optical mode. These proposals usually involve NOON states or other schemes with entanglement of the states with many photons. The interferometer being proposed by the author has a classical $\Delta\phi_{res} \propto \frac{1}{\sqrt{N}}$ scaling of the phase fluctuations with the number of photons in a laser mode but takes full advantage of the two-frequency configuration of the interferometer and consequent entanglement of channels. As is shown in Appendix, the scaling of the phase difference between channels can be approximately proportional to $\frac{1}{\sqrt{N_1 \cdot N_2}}$ .

Because this scheme does not depend on detection of the multiquantum non-classical states such as 'NOON' it is expected to be much more robust to external noise than the schemes based on achieving the Heisenberg limit (Ma et al., 2016).

## Experimental configuration

The interferometric setup consists of a powerful Earth-based quasi-continuous laser and three-four satellites in GSO within line of sight of the laser projector (see Fig. 1). Each satellite possesses a stablized nonlinear crystal inside a high-Q Fabri-Perot etalon (Fig. 2). [3] Unspecified but well-researched system of prisms and mirrors divides an original beam into three beams directed at satellites. [4] As a practical necessity, the laser setup must be placed on a high mountain to prevent significant attenuation of the beams by the Earth atmosphere but this author does not know whether it'll be sufficient on energetic grounds.

[1] Kean University (NJ), Wenzhou campus, China and SciTech Associates, LLC, University Park, PA. Contact e-mail: pblerner@syr.edu, plerner@kean.edu.

Spontaneous parametric down-conversion of the radiation incoming from the Earth-based laser is performed by a passive nonlinear element positioned on a geostationary satellite. For the purpose of discussion, we propose that the laser emits green light on the wavelength of the second harmonic of the Nd:YAG laser, $\lambda_L$=0.53 μ, which is down-converted into a red $\lambda_1$=0.88 μ and near-infrared photon $\lambda_2$=1.40 μ. Of course, the number of combinations of wavelength is infinite (for instance, $\lambda_1$=1.107 μ and $\lambda_2$=1.016 μ, both near to the $\lambda_{Nd:YAG}$=1.064 μ, for which there are well-developed amplifiers and other optical elements). Practical choice has to be defined by 1) transparency of the atmosphere, 2) properties of the nonlinear crystal and 3) existence of optical elements of sufficient quality for a given frequency. Further on, we will mention "green", "red" and "nIR" beams without going into detail though this identification for an actual wavelength is simply a matter of convenience.

**Principle of operation**

First set of coincidence detectors compares the difference between the red and nIR channels 1 and 2, respectively. Second set of coincidence detectors compares the difference between red$_2$ and nIR$_3$ and the third set of detectors compares the difference between red$_3$ and nIR$_1$. All three sets of the detectors are presumed to be locked on the null. The null position of the coincidence detectors corresponds to the following distances to the satellites $l_1$, $l_2$ and $l_3$:

$$(k \pm \tfrac{\Delta}{2})l_1 = (k \pm \tfrac{\Delta}{2})l_2 + \delta_1 + m_{12}\frac{\pi}{2}$$
$$(k \pm \tfrac{\Delta}{2})l_2 = (k \pm \tfrac{\Delta}{2})l_3 + \delta_2 + m_{23}\frac{\pi}{2} \qquad (1)$$
$$(k \pm \tfrac{\Delta}{2})l_3 = (k \pm \tfrac{\Delta}{2})l_1 + \delta_3 + m_{31}\frac{\pi}{2}$$

Here, k is the half of the green wavevector and Δ is half of the wavevector difference between wavectors of the red and nIR channels. Variables $\delta_1$, $\delta_2$ and $\delta_3$ are the phase shifts inside the interferometer itself emerging because of transmission and reflection. The multipliers m$_{12}$, m$_{23}$ and $m_{31}$ are insignificant integers. The system of Equations (1) contains five unknowns: $k, \Delta, l_1, l_2, l_3$. However, the entanglement of the paths provides three more equations only one of which is linearly independent from the first three:

$$\left(k + \frac{\Delta}{2}\right)l_1 = \left(k - \frac{\Delta}{2}\right)l_1 + \tilde{\delta}_1 + m_{11}\frac{\pi}{2}$$
$$\left(k + \frac{\Delta}{2}\right)l_2 = \left(k - \frac{\Delta}{2}\right)l_2 + \tilde{\delta}_2 + m_{22}\frac{\pi}{2} \quad \quad \quad (2)$$
$$\left(k + \frac{\Delta}{2}\right)l_3 = \left(k - \frac{\Delta}{2}\right)l_3 + \tilde{\delta}_3 + m_{33}\frac{\pi}{2}$$

An additional fifth equation in the case of three satellites is provided by the energy conservation for the parametric scattering:

$$k_L = 2k \quad \quad \quad \quad \quad \quad \quad \quad \quad \quad \quad \quad \quad \quad (3)$$

From these equations, five unknowns can be established exactly, in theory. In practice, there could be many sources of noise, instability of the pump laser frequency, in particular.

**Naive estimates of the resolution of the interferometer**

We consider that the green light power potentially achievable in each channel of the interferometer is 200 W, or, measured as flux, approximately $N_0 \approx 10^{21}$ light quanta per second. An effective divergence of the radiation with $D_1 = 10$ m mirror will be considered, with (upper) atmospheric turbulence and absorption being taken into account as $\Delta\theta_1 = 10^{-5}$ rad. So, we assume that the transmittance of the path allows about $2.5 \times 10^{-5}$ of the diffraction-limited intensity of the beam being imparted on the satellite's mirror:

$$\Delta\theta_D = \lambda/D_1 = 0.53 \cdot 10^{-6}\,m/10\,m = 0.53 \times 10^{-7} \quad \quad \quad (4)$$

Then, the incipient power on the satellite's own Fabri-Perot etalon with $D_2 = 1$m will be (for the distance of observatory from a geostationary satellite on the order of 40,000 km) $P_1 = 1.25$ mW. Assuming the recirculation of power in the high-fidelity Fabri-Perot etalon with $Q = 10^6$, the equivalent power for the production of the spontaneous photon pairs $P_2 = 1.25$ kW. The efficiency of the spontaneous two-photon parametric downconversion is estimated as $\rho = 10^{-9}$. This is a relatively high figure of spontaneous downconversion but the efficiency of only order of magnitude lower has been already achieved in a device intended for space applications. (Steinlechner, 2015) Of course, there is a long distance from achieving similar figures for the space installation, which is not intended to be optically adjusted and mechanically serviced for years. The downlinked power from a satellite is thus equal to 1.25 µW. By assuming an effective divergence on the downlink path at $\Delta\theta_2 = 10^{-4}$ rad, we arrive at approximately ~$5 \cdot 10^7$ "red" and nIR photons collected by the mirror similar in size to $D_1$. In practice, effective concentration of the space-based light can be much sharper than diffraction image as demonstrated by experiments of Janyi Yin and the light circle on the surface can be 5-15 m in diameter. [1] A phase shift observable in such an interferometer is estimated through a conventional formula. Yet, because of the entanglement of red and nIR paths between two satellites in each channel, a measured phase shift can achieve the value of (see the Appendix for the estimate)

$$< \Delta\phi^2 > \simeq \frac{1}{\sqrt{\overline{N_1} \cdot \overline{N_2}}} \approx \frac{1}{\overline{N}} \approx 2 \cdot 10^{-8} \tag{5}$$

Where $\overline{N}$ is an average photon number in each of the interferometer's arms. The phase shift of the Equation (5) corresponds to the measured displacement of $\Delta L = 5 \times 10^{-16}$ m. The additional gain in the sensitivity of the interferometer with respect to LIGO will be determined, similarly, to LISA, by the extremely long arms of the interferometer—L=$4 \cdot 10^7$ m. The sensitivity of the interferometer is expressed by

$$\frac{\Delta L}{L} = 1.25 \cdot 10^{-23} \frac{\sqrt{\Delta\nu_{Band}}}{\Delta\tau} \, sec. \tag{6}$$

Where $\Delta\nu_{Band}$—is the bandwidth, usually approximated by the inverse wavelength of the sensed wave in appropriate units ($\Delta\nu_{Band} \approx \frac{1}{7} Hz$) and $\Delta\tau$—is the time of signal accumulation by the detecting system. This author is not competent to decide whether such sensitivity is interesting for the cosmologists or not.

Appendix. **Phase sensitivity of the entanglement-locked interferometer**

We will estimate the phase sensitivity of the entanglement-locked interferometer by the naïve formula for the phase-difference operator:

$$\Delta\hat{\phi} = \left( \frac{\pi}{2\sqrt{\bar{N}+\Delta\hat{N_1}}} - \frac{\pi}{2\sqrt{\bar{N}+\Delta\hat{N_2}}} \right) \tag{A1}$$

In the Equation (A1), $\Delta\hat{N_1}, \Delta\hat{N_2}$ are the deviations of number operator in the modes 1 and 2 from an average number of photons $\hat{N}$. The number of photons is unknown but it is essential that red and nIR photons are coming in pairs. The expression (A1) looks rather absurd but the square roots of the positively-defined bound operators are also well defined. (Ref. [7] and *op. cit.*) While phase operator in optics is not well-defined, the phase difference operator has a definite meaning *mod*(2π). [11] Using Taylor expansion of the operators, which is also well defined, we arrive at the expression:

$$\sqrt{<\Delta\phi^2>} = \frac{\pi\sqrt{\langle(\Delta N_1 - \Delta N_2)^2\rangle}}{2\bar{N}^{3/2}} \tag{A2}$$

The formula for the average of two number operators ([7] and *op. cit.*) is as follows:

$$\left(\Delta\hat{N_1} - \Delta\hat{N_2}\right)^2 \geq \bar{N}(\bar{N}+1)cos^2\varphi + \bar{N}(\bar{N}+1)sin^2\varphi - 2Cov\left(\Delta\hat{N_1}, \Delta\hat{N_2}\right)sin\varphi cos\varphi \tag{A3}$$

Here, φ is the angle between two phase shifters. The maximum value for covariance of the two number operators is $N^2$. By adjusting the angle between two phase shifters to $\varphi = 45°$, we can completely compensate a term quadratic in the average photon number. The following equation results from (A2):

$$<\Delta\phi^2> = \frac{\pi}{4\bar{N}} \tag{A4}$$

Notably, the elimination of the $N^2$ term in (A3) by the phase rotation can be eliminated for the quantum efficiency of the detectors $\eta_{1,2} < 1$. Indeed, the equation (A3) acquires the form:

$$\left(\Delta\widehat{N}_1 - \Delta\widehat{N}_2\right)^2 >= \eta_1 \cdot \overline{N}(\eta_1 \cdot \overline{N} + 1)cos^2\varphi + \eta_2 \cdot \overline{N}(\eta_2 \cdot \overline{N} + 1)sin^2\varphi - 2\eta_1 \cdot \eta_2\overline{N}_1 \cdot \overline{N}_2 sin\varphi cos\varphi \qquad \text{(A5)}$$

and the required phase rotation between the detectors has to be $tg\varphi = \frac{\eta_2}{\eta_1}$.[2]

---

[2] The possibility of exact compensation of the shot noise was explained to the author by a genial, but untimely deceased V. N. Klyshko (1928-2000), the discoverer of the spontaneous parametric conversion of light.

**References**

1. J. Yin et al., *Science* 356, 1140 (2017)

2. D. Bouwmeester et al., *Nature* 390, 575 (1997)

3. B. Hensen et al., *Nature* 526, 682 (2015)

4. C.-Z. Peng et al., *Phys. Rev. Lett.* 94, 150501 (2005)

5. X.-S. Ma et al. *Nature* 489, 269 (2015)

6. T. Kim, M. Fiorentino and F. Wong, *Phys. Rev.* A 73, 012316 (2006)

7. Demkowicz-Dobrzanski, R. et al. Quantum limits in optical interferometry, arXiv:1405.7703v2 [quant-ph] 8 Oct 2014.

8. Yiqiu Ma et al. Proposal for Gravitational-Wave Detection Beyond the Standard Quantum Limit via EPR Entanglement, arXiv:1612.06934v1 [quant-ph] 21 Dec 2016.

9. M. H. A. Reck, Quantum Interferometry with Multiports: Entangled Photons in Optical Fibers, *Dissertation zur Erlangung des akademischen Gradeseines Doktors der Naturwissenschaften*, Juli 1996, University of Vienna.

10. Steinlechner, F. Sources of Photonic Entanglement for Application in Space, PhD Thesis, 2015, Institut de ciències fotòniques.

11. Loudon, R. *The Quantum Theory of Light, Oxford University Press*, Oxford: UK, 1973.

**Figure captions**

Fig. 1 Schematic drawing of the three-satellite constellation. A "green", i.e. high-frequency beam from the Earth-bound telescope undergoes spontaneous downconversion by the nonlinear elements (see Fig. 2) positioned at the three satellites on a geostationary orbit. Two beams from each satellite are collected by the observatory on Earth and analyzed by the detection scheme plotted on Fig. 3.

Fig. 2 Satellite-based nonlinear parametric converter. A "green" beam from Earth-based telescope is captured inside a high-Q Fabry-Perot resonator and is spontaneously downconverted by a parametric crystal shown as solid blue block into two lower frequency beams, shown by red and purple on the sketch. The axes of a nonlinear crystal are chosen so that two spontaneously downconverted photons propagate in the same direction.

Fig. 3 Detection scheme. Six beams from the satellites are divided by the six mirrors and the resulting pattern is observed by counting photons in each of the thirteen possible coincidence or anti-coincidence channels (see Fig. 4). Photo-detectors in the scheme are connected by thirteen logical gates, of which only six are independent. Semicircles indicated by $D_i$, i=1÷11 are the coincidence photon counters, circles indicate phase rotators, counters 12 and 13 are omitted for clarity of the picture. In the case of perfect quantum efficiency of the optical detection, $\varphi=45°$.

Fig. 4 The graph of the adjacency matrix of the optical detectors.

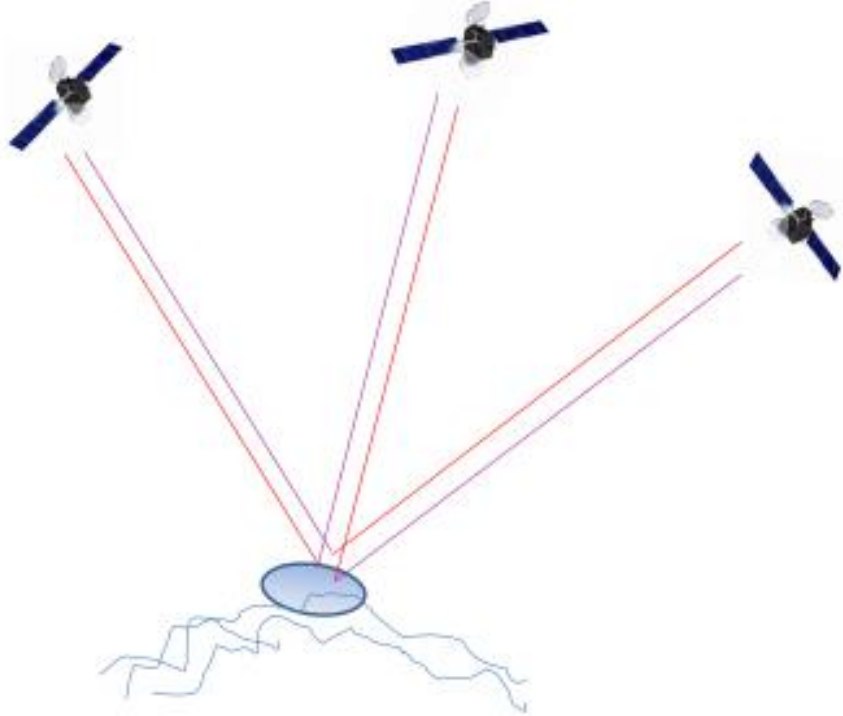

Fig. 1

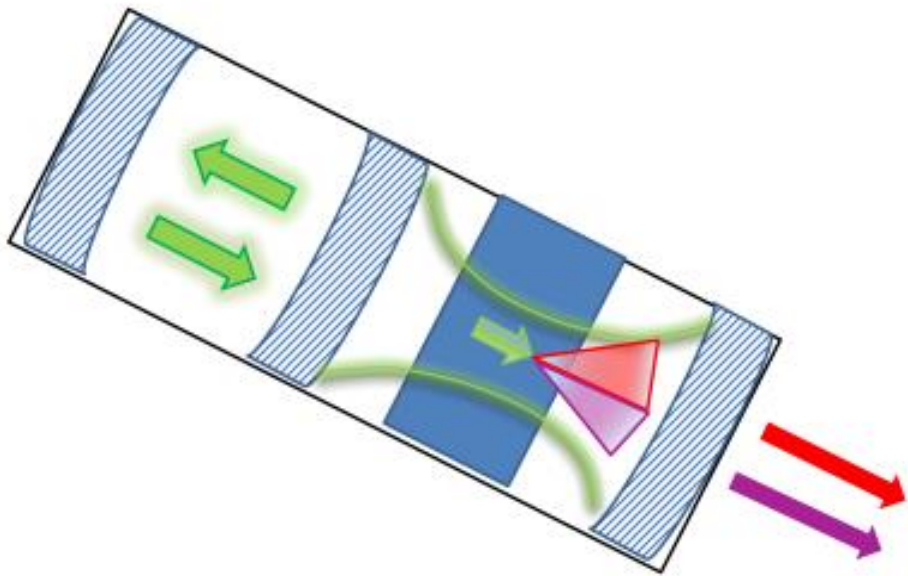

Fig. 2

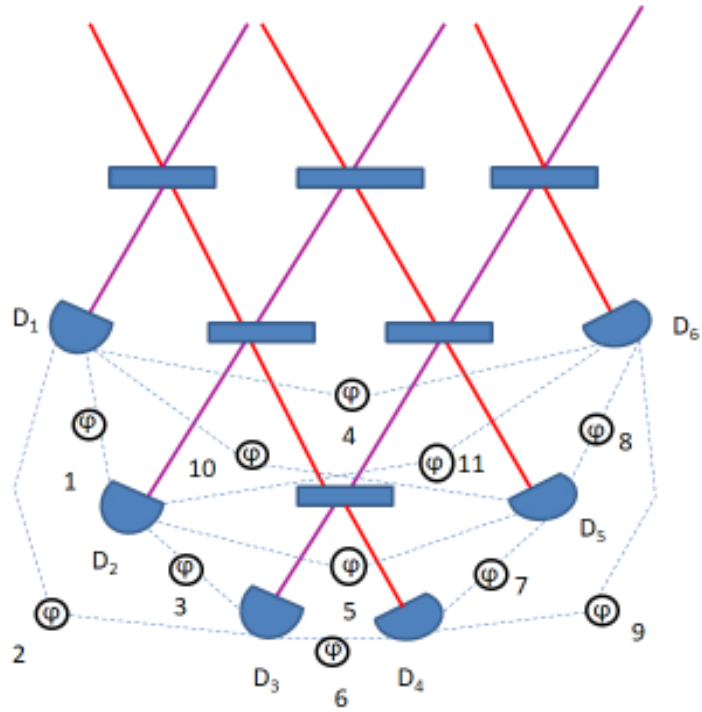

Fig. 3

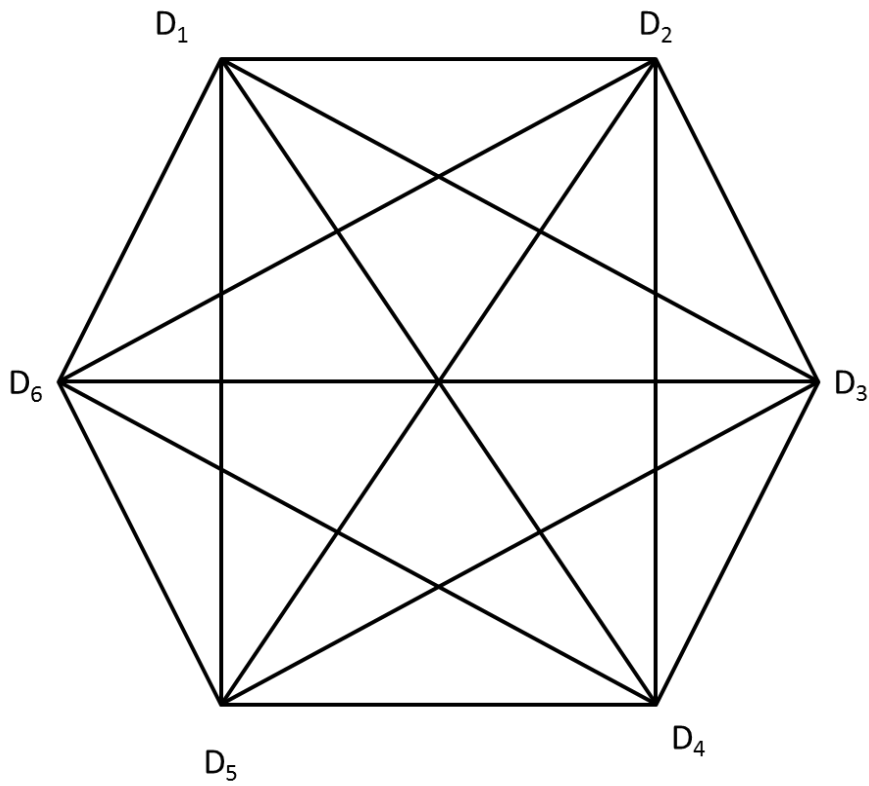

Fig. 4

9